\documentclass[12pt]{article}
\usepackage{graphicx}
\usepackage[latin1]{inputenc}
\usepackage{amsmath}
\usepackage{empheq}                       
\graphicspath{{./}{./figures/}}
\usepackage[a4paper, left=2.5cm, right=2.5cm, top=2cm, bottom=2.5cm]{geometry}

\usepackage{xcolor}
\usepackage[urlcolor=blue]{hyperref}      
\hypersetup{
    colorlinks = true,                    
    citecolor = {blue},
    linkcolor = {purple},
           }
\begin{document}
\title{Why planetary and exoplanetary protection differ: \\ The case of long 
duration Genesis missions to habitable but sterile M-dwarf oxygen planets}

\author{{\bf Claudius Gros}\\ \\
Institute for Theoretical Physics, \\
Goethe University Frankfurt, Germany} 

\maketitle

{\bf Keywords:} interstellar missions, directed energy launch systems,

magnetic sails, transiently habitable exoplanets, 

oxygen planet, Genesis project

\begin{abstract}
Time is arguably the key limiting factor for interstellar exploration. 
At high speeds, flyby missions to nearby stars by laser propelled 
wafersats taking 50-100 years would be feasible. Directed energy 
launch systems could accelerate on the other side also crafts weighing 
several tons to cruising speeds of the order of 1000\,km/s (c/300).
At these speeds, superconducting magnetic sails would be able to
decelerate the craft by transferring kinetic energy to the protons
of the interstellar medium. A tantalizing perspective, which would 
allow interstellar probes to stop whenever time is not a limiting 
factor. Prime candidates are in this respect Genesis probes,
that is missions aiming to offer terrestrial life new evolutionary 
pathways on potentially habitable but hitherto barren exoplanets. 

Genesis missions raise important ethical issues, in particular with 
regard to planetary protection. Here we argue that exoplanetary
and planetary protection differ qualitatively as a result
of the vastly different cruising times for payload delivering 
probes, which are of the order of millennia for interstellar
probes, but only of years for solar system bodies. Furthermore
we point out that our galaxy may harbor a large number of 
habitable exoplanets, M-dwarf planets, which could be sterile 
due to the presence of massive primordial oxygen atmospheres.
We believe that the prospect terrestrial life has in our galaxy
would shift on a fundamental level in case that the existence
of this type of habitable but sterile oxygen planets will be 
corroborated by future research. It may also explain why
our sun is not a M dwarf, the most common star type, but a 
medium-sized G-class star.
\end{abstract}

\section{Introduction}

Space exploration is confronted inherently with the extended travel times
needed to traverse the voids of space extending between earth and the
destination. In addition to travel times, key limiting factors are cost
considerations and the time devoted to mission development and design. The
latter is in particular the case for further away destinations, such as
missions to the outer solar system. Typically examples of past and future
missions are here the  Voyager crafts \cite{stone1977voyager}, the 
Europa Clipper \cite{phillips2014europa} and missions searching for 
life in the subglacial waters of ice moons \cite{konstantinidis2015lander}.
Is there however a maximal planning horizon societies would be willing 
to support? It is not uncommon for projects involving large-scale 
scientific or technology development tasks to span decades. Long-term 
collaborative efforts, like the ITER fusion reactor \cite{ikeda2009iter},
are nevertheless more often than not accompanied by continuous controversies
concerning the ultimate effort to utility ratio, with a central reason
being the rational to discount future rewards \cite{epper2011viewing}.
It is hence unlikely that explorative space missions taking centuries 
or even millennia to complete would ever survive the initial cost-to-benefit 
evaluation. The situation may however change for endeavors not designed 
for their usefulness in terms of science data or other return values. 
This will be the case, as we argue here, for Genesis missions aiming 
to establish an ecosphere of unicellular lifeforms on
potentially habitable but hitherto barren exoplanets.

\section{The utility vs. realizability dilemma of deep space exploration}

Solar system traveling times are long, but somewhat manageable. The recent
surge of interest in directed energy launching systems 
\cite{kulkarni2018relativistic} has presented us on the other side with 
the prospect that interstellar space missions may become realizable 
within several decades. The development of mission scenarios for
interstellar probes has hence left the realm of fusion-size 
spaceships \cite{freitas1980self,long2011project}, becoming instead a 
question of pros and cons. We argue here that two types of interstellar 
probes may be considered, fast data return probes and
comparatively slow Genesis crafts.

\subsection{Data gathering by fast flyby probes}

Directed energy launch systems may accelerate wafersat spacecrafts 
weighing a few gram up to 20\% of the speed of light 
\cite{brashears2016building}, at least modulo a substantial technology 
development effort. Wafer-sized interstellar probes capable of
reaching the nearest stars within several decades may hence be 
employed, as envisioned by the Breakthrough Starshot Initiative 
\cite{parkin2018breakthrough}, for science data return
missions. The technical challenges, ranging from material issues 
\cite{atwater2018materials}, the stability at launch 
\cite{srinivasan2016stability}, to the interaction of a 
relativistic spacecraft with electromagnetic forces 
\cite{hoang2017electromagnetic}, and with the interstellar medium 
\cite{hoang2017interaction}, are immense,
but not insurmountable.

Interstellar science probes need to be fast, as data 
return missions taking centuries would not be considered 
worth the investment. Flyby missions are consequently the 
only viable option. Decelerating with the aim to enter solar
or planetary orbits involves on a practical level the 
transfer of the kinetic energy to either the photons of the 
target star or to the ionized particles of the interstellar
medium. Solar sail deceleration, the first method, is however 
not possible when the craft travels at relativistic speeds; 
the probe would have bypassed the target star long before 
coming to a stop. For the second method, magnetic sails
weighing of the order of several hundreds tons would be 
necessary \cite{gros2017universal}. To accelerate a craft 
of that size to close to the speed of light is however 
beyond launch infrastructures potentially realizable within
the next generations. The same holds for active deceleration 
of massive spaceships by TW-sized solar-system based laser 
beams \cite{forward1984roundtrip}. Interstellar science 
probes are consequently realizable only if they are fast, 
viz when limited to flyby investigations.

\subsection{Payload delivery by slow interstellar crafts}
Fast and slow are highly relative terms in the realm of space 
travel. The Voyager spaceprobes cruise at speeds of the order 
of 20\,km/s, which is high with respect to everyday's velocities, 
but slow when it comes to interstellar distances. Here we 
consider an interstellar spacecraft to be `slow' when cruising 
roughly 50 times faster than the Voyager probes, at 1000\,km/s, 
which corresponds to 1/300 of the speed of light. The millennia 
needed to reach the nearest stars at such a velocity exceed 
typical human planing horizons by far, which implies that it is 
not possible to assess the potential benefits of slow interstellar 
missions with standard utility criteria. A slow craft has on the 
other hand enough time at its disposal to decelerate via magnetic
braking, that is by transferring momenta to the interstellar protons
\cite{perakis2016combining}. It has been estimated in this regard, that 
the magnetic field created by a 1.5\,ton superconducting loop with a 
radius of 50\,km would be able of doing the job \cite{gros2017universal}.
Magnetic sails are moreover self-deploying, as wires with opposite
currents repel each other. Slow interstellar spacecrafts are hence 
suited for delivering ton-sized payloads to far away destinations. 

Slow interstellar ton-sized crafts may be launched, importantly, by the same
directed energy launch systems envisioned for fast flyby missions, with the
reduced velocity trading off the increased weight \cite{lubin2016roadmap}. 
Comparatively slow, that is non-relativistic interstellar crafts, could 
be accelerated alternatively by the type of advanced ion engines that
are being developed within NASA's evolutionary Xenon thruster (NEXT) 
effort \cite{soulas2004next,shastry2015status}.
Laser arrays of the order of 100\,MW would be used in this case not to 
propel a reflecting light-sail, but to power the solar cells of the 
craft \cite{brophy2018breakthrough}. A conversion rate of 70\%, potentially 
achievable when the sail performance 
is tuned to the laser frequency, would then be enough to power 70\,MW 
lithium-fueled gridded ion thrusters \cite{brophy2017breakthrough}. 

Overall we are confident that interstellar spaceprobes entering 
solar or planetary orbit on arrival are potentially realizable, 
albeit at the cost of prolonged mission times. Less likely seem
in contrast the perspective that fast probes could decelerate, 
independently of the technique envisioned for the 
braking maneuver.

\subsection{Interstellar deceleration}

The lack of an in-place infrastructure implies, on arrival,
that it is substantially more challenging to decelerate an 
interstellar craft than to speed it up in first place. This
is in especially true when the craft is fast and when mission
durations should be kept within human planing horizons.
Given enough time, and  a magnetic sail of substantial size
\cite{gros2017universal,zubrin1991magnetic,freeland2015_mathematics}, 
braking from the protons of the interstellar medium is however 
feasible, as discussed in the previous sections.

Regarding solar sails, it has been proposed that graphene may 
be the optimal candidate material \cite{matloff2012graphene},
both for launching an interstellar probe and for braking from 
the photons of the target star \cite{heller2017deceleration}. 
Physically, a graphene monolayer is characterized by an ultra-low  
areal mass density of $7.4\cdot10^{-7}\,\mathrm{kg}/\mathrm{m}^2$, 
a negligible reflectivity and a flat absorption coefficient 
$A_\omega=\pi\alpha\approx 0.023$, as resulting from the 
Dirac cone, where $\alpha= e^2/(\hbar c) \approx 1/137$
is the fine structure constant \cite{nair2008fine}. Assuming 
that the properties of a graphene monolayer could be improved 
by about a factor 100 without a corresponding weight increase,
namely to reflectivity values of 99.99\%\,--\,99.999\%,
a fast interstellar craft could decelerate at $\alpha$-Centauri
using stellar photon pressure \cite{heller2017deceleration},
at least as a matter of principle. How to realize the required 
performance boost is presently however unclear.

The field of a magnetic sail is produced by a large superconducting
loop. Alternatively one may consider an electric sail 
\cite{janhunen2007simulation}, which consists of electrically charged
structures of similar extensions. In this case it is the electric
field of the charged craft that reflects the protons of the 
interstellar medium \cite{perakis2016combining}. It would however
be a challenge for an interstellar craft to power the electron
gun needed to maintain the required potential difference between
the craft and the surrounding rarefied medium. Electric sails may however
be advantageous for solar system application, as their performance
decays only as $1/r$, as a function of the distance to the sun,
and not as $1/r^2$, as for solar sails \cite{wiegmann2018summary}.

\subsection{The slow path from prokaryotes to eukaryotes}

Payload delivering interstellar crafts come with cruising times of a 
few millennia, at least, that is with timescales that may seem
extraordinary long to human planning horizons. A handful of millenia
are on the other hand irrelevant from the perspective of evolutionary 
processes. On earth it took about one billion years, that
is until the end of the archean genetic expansion \cite{david2011rapid}, 
to develop modern prokaryotes, viz bacteria, and another billion years 
for the basis of all complex life, eukaryotic cells, to emerge 
\cite{parfrey2011estimating}. It is not a coincidence, that higher
life forms are made of eukaryotic and not of prokaryotic cells,
but a consequence of the energy barrier that prevents prokaryotic 
cells to support genomes of eukaryotic size \cite{lane2010energetics}.
The massive genomes necessary for the coding of complex eukaryotic
morphologies are typically four to six orders of magnitude larger 
than the genetic information encoding prokaryotic life \cite{lane2014bioenergetic}.

The emergence of eukaryotic cells has been on earth the key bottleneck
along the route from uni-cellular to multi-cellular and morphological
complex life. Taking the timescale of terrestrial evolutionary processes 
as a reference, we may hence postulate that exobiological lifeforms could 
need similar time spans, if at all, to evolve to complexity levels comparable 
to terrestrial eukaryotic life \cite{cockell2015astrobiology}. 
There are moreover arguments for the possibility, as discussed further 
below, that a relatively high percentage of potentially habitable planets 
may harbor either no life at all, or only lifeforms equivalent in 
complexity below that of modern terrestrial bacteria. A payload of 
single-cell eukaryotes, either as germs or in terms of codings for an
onboard gen laboratory, would hence be a valuable payload for slow 
cruising Genesis probes destined to habitable but otherwise barren 
exoplanets. Operationally, instead of landing, the Genesis probe would 
carry out the seeding process from orbit via the retrograde expulsion 
of micro-sized drop capsules. The goal would be in the end to lay the 
foundations for a self-developing ecosphere of initially unicellular 
organisms \cite{gros2016developing}. 

\section{The case for habitable but sterile oxygen planets}

A Genesis probe should comply with planetary protection considerations 
and target only certain types of potentially habitable planets
\cite{frick2014overview}. One possibility is that the candidate planet is only 
transiently habitable, that is for time spans that are too short for 
complex single- or multi-cellular life to develop \cite{gros2016developing}.
Examples for causes for limited habitability are orbital instabilities of
the hosting planetary system and geological disruptions due to the 
absence of plate tectonics \cite{gros2016developing}. Of interest
in this respect are furthermore planets orbiting brown dwarfs 
\cite{bolmont2018habitability}, that is failed stars having 13-75 
times the mass of Jupiter. The mass of brown
dwarfs is too low for hydrogen fusion, the energy source of main-sequence
stars, with the consequence that the star cools progressively by radiative
dissipation of its initial reservoir of thermal energy. Depending on the 
mass of the star, on the orbital distances of the planets and on other 
parameters, like the impact of gravitational and atmospheric tides, a 
given brown-dwarf planet could remain habitable for periods ranging 
from a few hundred million years to a few billion years \cite{bolmont2018habitability}. 
Brown dwarf planets are hence interesting Genesis candidate planets.

\subsection{Abiotic oxygen buildup in the runaway greenhouse state of young M-dwarf planets}

Stars with a mass greater than about 0.075 the mass of the sun are heavy enough
to produce energy via hydrogen burning. A well known example is the Trappist-1
system \cite{gillon2017seven}, a system composed of seven earth-sized planets 
orbiting a M-dwarf star at distances that are either within or close to the 
nominal habitable zone. The mass of the central star is in this case about 0.08 
the mass of the sun, which is not a coincidence. Estimates show 
\cite{dressing2015occurrence}, that a majority of rocky habitable zone 
planets is expected to orbit M dwarfs, that is low-mass stars
like Trappist-1.

M dwarfs are characterized by an extended Kelvin Helmholtz contraction 
time, which is the time it takes for a protostar to reach the main 
sequence by shedding its initial reservoir of gravitational energy
radiatively. The Kelvin Helmholtz timescale extends from about 10 million years
for sun-like stars to several hundred million years for late M dwarfs 
\cite{stahler2008formation}.
Planets orbiting low mass stars at a distance corresponding to the
main-sequence habitable zone will hence experience an extended initial runaway
greenhouse state induced by the increased irradiation from the initially
substantially larger host star. With the ending of the Kelvin Helmholtz
contraction of the central star the atmosphere of the planet cools
correspondingly. 

In the initial greenhouse state the stratosphere is wet. The UV radiation of
the host star leads in this stage to the photodissociation of water, and 
with it to the loss of hydrogen to space, with the oxygen staying mostly behind
\cite{luger2015extreme}. Depending on the initial reservoir, several earth 
oceans worth of water may be lost altogether \cite{tian2015water}. For the 
habitable-zone planets of the Trappist-1 system the resulting buildup of 
abiotic oxygen has been estimated to reach partial pressures of 350-490 bars 
\cite{bolmont2016water}. It is presently not clear to which extent the 
buildup of abiotic oxygen during the Greenhouse state is countered by redox 
reactions resulting from the interaction of the atmosphere with a magma 
ocean \cite{wordsworth2018redox}. It is likely that the final oxygen content 
of the atmosphere is reduced, but still substantial. Primordial oxygen partial
pressures of several bars and more may hence be a common feature of rocky
M-dwarf planets.

\subsection{Are oxygen planets sterile?}
The chemical environments of oxygen planets, that is of planets 
disposing of a substantial amount of primordial atmospheric oxygen,  
are expected to differ substantially from the one of archean earth. 
The origins of life on earth are yet not understood \cite{tor2017catalyst}, 
it is however clear that abiogenesis may occur only in microstructured
chemo-physical reaction environments \cite{herschy2014origin} that
are driven by a sustained energy source \cite{russell2010serpentinization},
as realized within the alkaline hydrothermal vent scenario 
\cite{sousa2013early}. Potential birthing places of life such
as submarine alkaline vents are conjectured furthermore to be characterized 
by steep electrochemical concentration gradients \cite{russell2013inevitable},
as a necessary precondition for the emergence of prebiotic vectorial reaction 
pathways. Primordial oxygen, when present, could disrupt however
the formation of these electronchemical disequilibria \cite{sojo2016origin}. 
An important point in this context is a well-known relationship between 
oxygen and cellular energy\footnote{W.\,F.~Martin, private
communication}, namely that the synthesis of the chemical constituents of 
cells, like amino acids, bases and lipids, from glucose and ammonium,
demands about 13 times more energy per cell in the presence of O$_2$ 
than in the absence of oxygen \cite{mccollom2005thermodynamic,lever2015life}.
It is hence conceivable that the emergence of life could be preempted on 
otherwise habitable M-dwarf planets by the presence of primordial oxygen. 
A substantial amount of future research effort is clearly warranted in
order to corroborate, or to disprove this presumption. In case, we 
would live in a galaxy where habitable but sterile planets abound. 
Oxygen planets would then be prime candidates for Genesis missions.

\section{Planetary vs.\ exoplanetary protection}

An endeavor aiming to endow other planets with life raises a 
series of ethical issues. From a utilitarian perspective it 
may be considered in fact unethical to allocate a substantial 
amount of resources to projects not contributing to
the overall welfare of humanity \cite{sheskin2016switching}.
We will not pursue this argument further, focusing
instead on two key aspects of planetary protection.

\subsection{Planetary protection for human benefit}
Planetary protection had been formulated historically with the 
exploration of the solar system in mind \cite{devincenzi1984revised}. 
Back contamination needs to be avoided, clearly, such that {\em ``earth 
is protected from the potential hazard posed by extraterrestrial matter 
carried by a spacecraft returning from another planet''}. Space exploration 
should be carried out, furthermore, in a manner that does not
jeopardize {\em ``the conduct of scientific investigations of possible
extraterrestrial life forms, precursors and remnants''}. Human benefit
considerations have hence been, as these formulations of the planetary
protection policy of the International Committee on Space Research (COSPAR)
show, the core original rational for avoiding not only backward, but also
forward contamination \cite{sherwood2018forward}.

The very existence of extraterrestrial life is a subject of debate. 
Remote sensing attempts, like the detection of extrasolar life via 
a direct or indirect spectral analysis of exoplanetary atmospheres 
\cite{seager2014future}, will be carried out in the next years. 
In situ investigations of extrasolar life are in contrast unlikely to
be ever undertaken. On one side because the delivery of the required 
landing modules by slow-cruising interstellar probes would take millennia. 
The second point is that we may expect science to progress within the 
intervening centuries to a point that would allow for a near to full 
understanding of the possible routes to abiogenesis and of the spectra 
of possible lifeforms. Another aspect is that computer experiments can
be anticipated to advance to a point that would allow, eventually, to 
retrace the geophysical evolution of a given non-solar planetary system 
in detail, possibly when supplemented by flyby observations. Relatively 
little could be added in this case by additional in situ investigations. 
Protecting the rudimentary biosphere of an exoplanet for science purposes 
is hence not as relevant as it is for solar system bodies.

\subsection{Ethically grounded planetary protection}
Common ethical imperatives are ambiguous when human activities 
impact higher but non-human life forms, in particular with regard 
of the relative relevance of anthropocentric and non-anthropocentric 
values \cite{callicott1984non}. There is however a deeply rooted 
common-sense notion that humanity should protect life forms of a
certain level of complexity, at least whenever possible. This notion 
withstands the Darwinian nihilist viewpoint \cite{snyder2017darwinian}, 
attributing instead value to life per se \cite{persson2017ethics}. 

Taking the evolution of terrestrial biota as a reference 
\cite{david2011rapid,gros2016developing}, we may
classify non-solar ecosystems into four categories: primitive-prokaryotic,
prokaryotic, unicellular eukaryotic and multi-cellular eukaryotic, viz complex
life. For terrestrial life it is custom to attribute value nearly exclusively
to complex life, viz to animals and plants. Killing a few billion bacteria
while brushing teeth does not cause, to give an example, moral headaches. The
situation changes however when it comes to extrasolar life, for which we may
attribute value also to future evolutionary pathways. This is a delicate
situation. Is it admissible to bring eukaryotes to a planet in a prokaryotic
state, superseding such indigenous life with lifeforms having the potential
to develop into complex ecologies? Our prevalence to attribute value
predominately to complex lifeforms would suggest that this would be 
ethically correct \cite{snyder2017darwinian,persson2017ethics}, in particular 
if we could expect our galaxy to harbor large numbers of planets in prokaryotic 
states. Endowing a selected number of exoplanets with the possibility to 
evolve higher life forms would in this case not interfere
with the evolution of yet simple life forms on potentially billions of other
planets.

Genesis missions would comply with the common-sense norm to 
attribute value to complex lifeforms, the very rational to 
undertake them in first place, and abort whenever the target planet
harbors life that can be detected from orbit. Considering the 
case of Mars, it is however clear that it will be hard to rule 
out unambiguously the existence of ecospheres of exceedingly low bioproductivity. 
Protocols regulating the necessary level of confidence are hence 
needed. It would be meaningful to embargo the entire extrasolar 
system in case that complex life would be detected by flyby probes 
on one of its planets.

\section{Discussion \& outlook}

The recent advent of directed energy launch concepts demonstrates 
that interstellar space probes may become realizable within the 
foreseeable future \cite{bible2013relativistic}. The technical challenges 
involved are daunting. An example is the development of self-healing 
electronics \cite{moon2016sustainable}, that is of circuits that would be
capable to withstand decades to millennia of cosmic bombardment 
\cite{stevenson2017self}. It is hence important to assess and to classify 
the range of possible interstellar missions. The first option is a high 
speed flyby mission by gram-sized wafersats that have been accelerated 
to a sizable fraction of the speed of light \cite{lubin2016roadmap}, 
say 20\%. Here we have pointed out that the directed energy launch
systems envisioned for fast flyby missions would be suited to launch in
addition payload delivering probes cruising at reduced velocities of typical
1000\,km/s. These probes would weigh of the order of several tons, in particular
due to the weight demands of the magnetic sail that would needed for braking
off the interstellar medium \cite{gros2017universal}.
The long arrival times of a minimum of several
thousand years require however an in depth analysis of the rational for
carrying out this kind of comparatively slow-cruising interstellar missions.
One possibility would be the Genesis project \cite{gros2016developing}, 
which proposes to initiate the development of precambrian ecospheres of 
unicellular organisms on transiently habitable exoplanets.

We have pointed out, in addition, that the existence of habitable 
but sterile oxygen planets would alter radically our view of our 
cosmic neighborhood, in particular from the perspective of interstellar 
mission planing. The number of potentially habitable M-dwarf planets 
has been estimated to be substantial \cite{wandel2018biohabitability}, 
with the consequence that it is not implausible that a rich biosphere
might be detected eventually on a nearby M-dwarf planet via remote sensing.
Biosphere compatibility considerations suggest in this case that we should 
not consider in-situ investigations of exoplanets teeming with life 
\cite{gros2016developing}, with the reason being that such an endeavor 
could be catastrophic for the indigenous biosphere. 

The situation changes, in contrast, if the target habitable planet 
contains a substantial amount of primordial atmospheric oxygen and if
primordial oxygen preempts the emergence of life. Habitable oxygen
planets would then be sterile. Oxygen, which is otherwise a 
preconditions for multi-cellular and hence complex life to thrive, 
is expected to be generated in vast amounts during the the initial 
runaway greenhouse state occurring during the extended Kevin Helmholtz 
contraction phase of nominally habitable late M dwarfs planets
\cite{tian2015water}. It is presently not known if the resulting 
primordial oxygen atmosphere, which may differ drastically from planet 
to planet in volume \cite{wordsworth2018redox}, would inhibit life to
originate in first place. The existence of habitable but sterile oxygen 
planets, that is of worlds that would offer terrestrial life nearly 
unlimited grounds for the pursue of new evolutionary pathways, would
revolutionize in any case our view of our galactic neighborhood. 

The initial Kevin Helmholtz contractions phase of yellow G-class 
stars like our sun is relatively short, typically of the order of 
several million years. Potentially habitable planets orbiting not 
a M dwarf, but G stars, are hence not forced to go through an extended 
initial Greenhouse state, even though they can enter one, like Venus, 
as a consequence of the final orbital parameters. One may speculate
whether this circumstance is the reason why earth is not orbiting a 
red M dwarf, the most frequent star type of the galaxy, but a star
type which is substantially less common, a yellow G star.

Regarding the difference between the protection of solar
system bodies and exoplanets we have pointed out that the 
extended time scales necessary for an in-situ exploration
of exoplanets changes the rational. Financing a deep-space mission 
taking several millenia cannot be justified along the lines
of solar system exploration, viz for the advancement of science.
It is interesting in this context to connect to the ongoing
controversy \cite{fairen2017searching}, whether the hypothetical 
counterfactual of planetary protection, {\em ``You protect what 
you want to study, but you cannot study what you protect.''}, does 
impede the search for life on Mars \cite{rummel2017four}. 
Protecting life on exoplanets for the sake of science is
in analogy not a valid rational, as it could be studied in
any case only on time scales far exceeding standard human planning 
horizons.

\section{Acknowledgments}

The author thanks 
Andr\'es de la Escosura Navazo,
Phil Holliger,
Kepa Ruiz Mirazo, Steven Beckwith,
William F.\ Martin and 
Michael J.\ Russell
for exchanges regarding the likelihood
of life emerging in the presence of oxygen.

\bibliographystyle{unsrt}

\end{document}